\documentclass[11pt,twoside]{article}
\usepackage{asp2010}

\resetcounters

\begin{document}

\title{Magnetic Field Decay Makes Neutron Stars Look Older Than They Are}
\author{Shuang-Nan Zhang $^{1,2}$ and Yi Xie $^1$
\affil{$^1$National Astronomical Observatories, Chinese Academy Of
Sciences, Beijing 100012, China.} \affil{$^2$Key Laboratory of
Particle Astrophysics, Institute of High Energy Physics, Chinese
Academy of Sciences, Beijing 100049, China.}}

\begin{abstract}
It is commonly accepted that a neutron star is produced, when a
massive star exhausts its nuclear fuel and ends its life in a
core-collapse supernova explosion. This scenario is confirmed by the
detection of pulsars, which are believed to be rapidly spinning
neutron stars, in the central regions of many supernova remnants.
Neutron stars and their associated supernova remnants should
therefore have the same ages. As expected, the age of the Crab
Pulsar, the first to be connected with a supernova remnant (the Crab
Nebula), can be inferred from its current spin period and its
derivative and indeed has about the same age of the supernova
remnant, that was produced from a historically recorded supernova
explosion in 1054. However most neutron stars appear to be much
older than the ages of their associated supernova remnants, a puzzle
not yet understood. Another puzzle is that so far no convincing
evidence has been found in favor of magnetic field decay in neutron
stars, that is predicted in most models of neutron stars. Here we
show convincing evidence of magnetic field decay in some young
neutron stars, and that the magnetic field decay can alter their
spinning behaviors significantly such that these neutron stars
appear much older than they really are.

\end{abstract}

\section{Magnetic Field Decay of Neutron Stars}

A rapidly spinning neutron star loses its rotational energy by
magnetic dipole radiation (Pacini 1968) and outflow of charged
particles accelerated in its magnetosphere (Goldreich 1969), i.e.,
$\dot{\omega}=-K \omega^3/(2\pi)^2$, where $\omega$ is its angular
frequency and $K\propto M^2/I$, in which $M$ is its dipole magnetic
moment and $I$ is its moment of inertia; its surface dipole magnetic
field strength can thus be estimated as, $B=3.3\times
10^{19}(P\dot{P})^{1/2}$, where $P$ and $\dot{P}$ are its spin
period and period derivative, respectively. Then assuming
$\dot{K}=0$, the neutron star's age can be calculated as, $T_{\rm
s}=(P-P_{0})/2\dot{P}$, or if $P\gg P_{0}$, $T_{\rm s}=P/2\dot{P}$,
commonly referred to as its characteristic or spin-down age. More
generally, the spin-down law can be written as,
$\dot{\omega}=-K\omega^n(2\pi)^{1-n}$, or, $\dot{P}=KP^{2-n}$.
Defining a measurable quantity, the so-called braking index, $n_{\rm
b}=2-P\ddot{P}/\dot{P}^2$, it is easy to show (Lyne 2006), $n_{\rm
b}=n-\dot{K}P^{3-n}/\dot{P}$. Assuming $n=3$, $n_{\rm b}<3$
indicates $\dot{K}>0$, or {\it vice versa} (Lyne 1975, Chanmugam
1995, Lyne 2004, Lyne 2006, Chen 2006). However the number of
neutron stars with reliably measured $n_{\rm b}$ is currently very
small (Chen 2006).

Many studies on the possible magnetic field decay of neutron stars
have been done previously, based on the observed statistics of $P$
and $\dot{P}$ of neutron stars. Some of these previous studies used
the spin-down ages as indicators of the true ages of neutron stars
and found evidence for their dipole magnetic field decay (Pacini
1969, Ostriker 1969, Gunn 1970); however as we will show in the
following, their spin-down ages are normally significantly different
from their true ages. Alternatively other studies relied on
population synthesis that includes complicated observational
selection effects, unknown initial parameters of neutron stars, and
not completely understood models of radio emission of pulsars (Holt
1970, Bhattacharya 1992, Han 1997, Regimbau 2001, Gonthier 2002,
Guseinov 2004, Aguilera 2008, Popov 2010). Consequently so far there
is no consensus on if and how neutron star's magnetic field decays
(see (Harding 2006a, Ridley 2010, Lorimer 2011) for reviews).

The true age of a neutron star can be given from the age of the
supernova remnant physically associated with it, since they both
were produced during the same supernova explosion. A supernova
remnant expands and interacts with its surrounding interstellar
medium, and thus its age, $T_{\rm SNR}$, may be calculated by
modeling its morphology at the current epoch. The ages of neutron
stars obtained this way are independent of the properties of neutron
stars and thus unbiased in principle. In figure~1~(upper panel) we
show the distribution of $T_{\rm SNR}/T_{\rm s}$; clearly for many
of them $T_{\rm SNR}/T_{\rm s}\ll 1$, i.e., many neutron stars look
much older than their associated supernova remnants. The
distribution is possibly bi-modal, divided by $T_{\rm SNR}/T_{\rm
s}\approx0.5$ ; we will return to this point later. In
figure~1~(lower panel) we show the correlation between $T_{\rm
SNR}/T_{\rm s}$ and the dipole magnetic field strength $B$. The
strong and significant positive correlation suggests that low values
of $B$ at the current epoch is the major cause for $T_{\rm
SNR}/T_{\rm s}\ll 1$. This immediately suggests that magnetic field
decay plays a significant role in making neutron stars appear older
than they really are, at least for neutron stars younger than about
one million years. This general idea was suggested previously (Lyne
1975, Geppert 1999, Ruderman 2005), but so far has not been compared
between data and models statistically.

\begin{figure}
\centerline{ \hbox{ \epsfxsize=3.0in \epsfbox{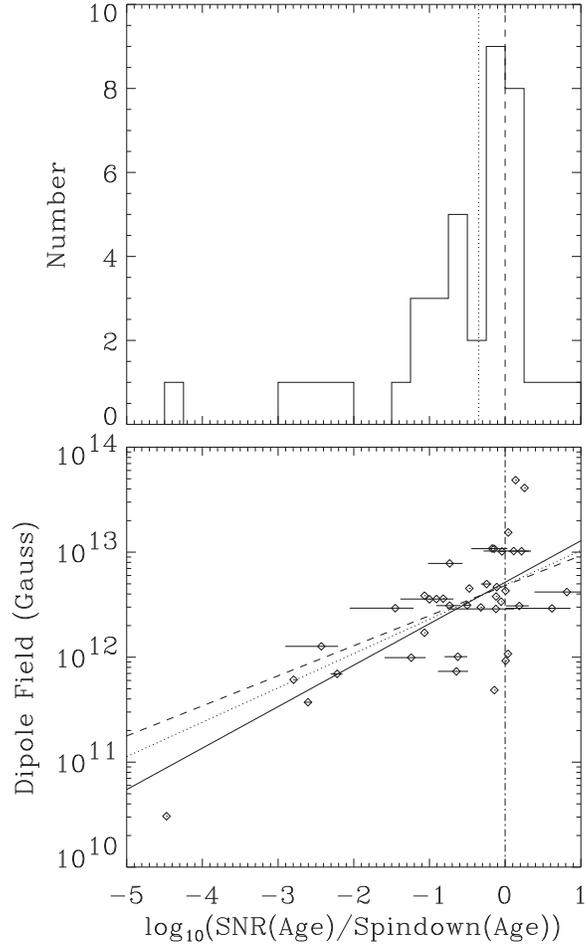}}} \caption{{\it Upper Panel}: Distribution of the
ratio between the age of a supernova remnant ($T_{\rm SNR}$) and the spin-down age ($T_{\rm s}$) of the
associated neutron star. The vertical dashed and dotted lines correspond to $T_{\rm SNR}=T_{\rm s}$ and $T_{\rm
SNR}=0.5T_{\rm s}$, respectively. {\it Lower Panel}: Correlation between the ratio and the dipole magnetic field
strength. The solid, dotted, and dashed lines correspond to linear fitting results (in double logarithm scales)
with all data points, the left-most point removed, and the five left-most points removed, respectively. In each
case a strong and significant positive correlation is found.}
\end{figure}

With the true ages of neutron stars available, we can then study the
correlations between their ages and their other two observables
normally available with each pulsar, namely, $P$ and $\dot{P}$, as
shown in figure~2. Assuming heuristically that all neutron stars
were born with the same properties, the trends shown here can be
explained as the evolutionary tracks of neutron stars. These trends
suggest that, as neutron stars become older, their periods become
longer, their period derivatives becomes smaller, and their dipole
magnetic field becomes weaker. One may argue that the observed trend
of their dipole magnetic fields may be a mathematical artifact,
because $B$ and $\dot{P}$ are positively correlated mathematically.
However this argument does not hold, because $B$ and ${P}$ are also
positively correlated in the same way mathematically. It is actually
physically more meaningful to conclude that it is the magnetic field
decay of neutron stars that causes $\dot{P}$ decreases rapidly with
time.

\begin{figure}
\centerline{ \hbox{ \epsfxsize=3.0in \epsfbox{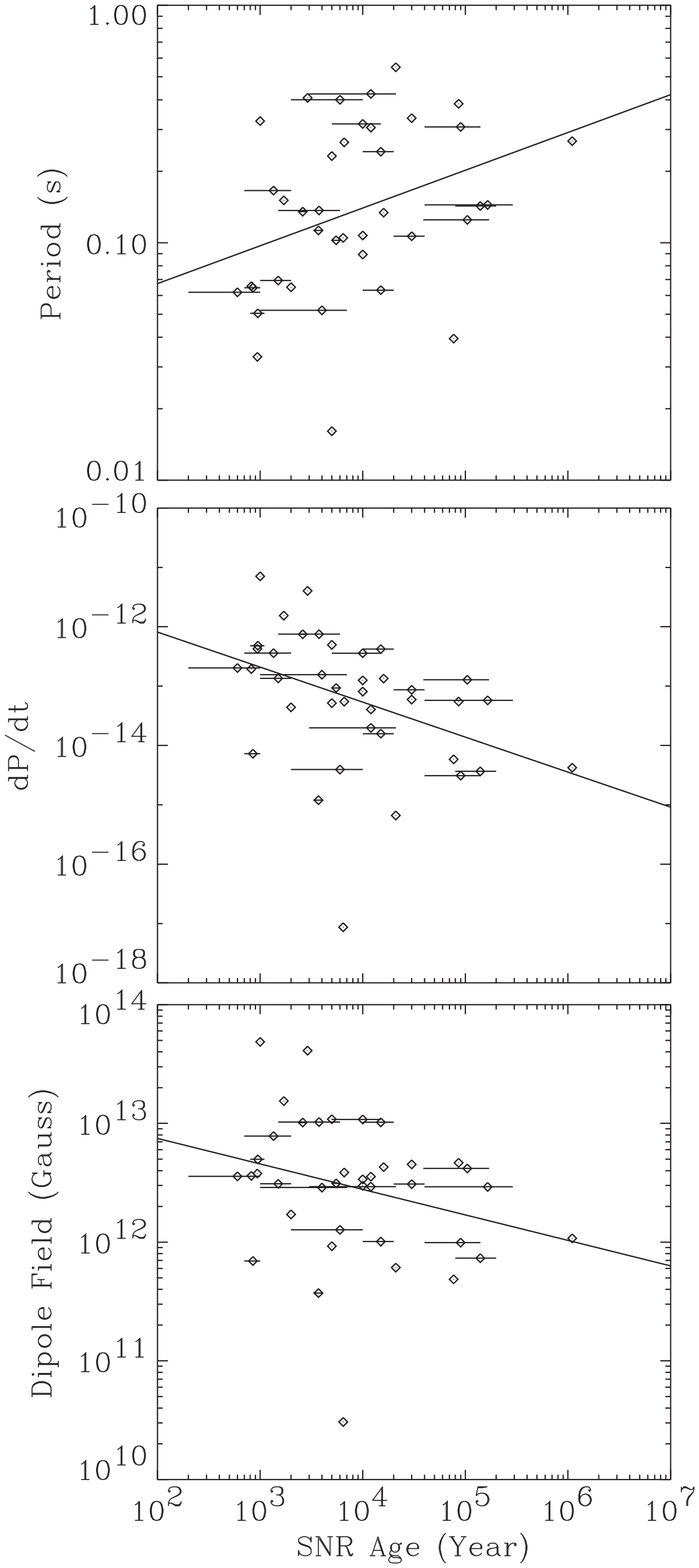}}} \caption{Correlations between the true ages ($T_{\rm
SNR}$) of neutron stars and their periods ($P$), period derivatives ($\dot{P}$), and surface dipole magnetic
field strengths ($B=3.3\times 10^{19}(P\dot{P})^{1/2}$). The straight lines are linear fitting results (in
double logarithm scales).}
\end{figure}

Several physical mechanisms have been proposed for magnetic field
decay in neutron stars, e.g., ohmic dissipation, Hall effect, and
ambipolar diffusion. We can therefore have (Goldreich 1992, Heyl
1998), $\frac{dB}{dt} = -B(\frac{1}{t_{\rm ohmic}}+\frac{1}{t_{\rm
ambip}}+\frac{1}{t_{\rm hall}})$, in which $t_{\rm
ohmic}\sim2\times10^{11}\frac{L_5^2}{T_8^2}(\frac{\rho}{\rho_{\rm
nuc}})^3~{\rm yr}$, $t_{\rm ambip}^{\rm
s}\sim3\times10^{9}\frac{L_5^2T_8^2}{B_{12}^2}~{\rm yr}$ (for the
solenoidal component), $t_{\rm ambip}^{\rm
ir}\sim\frac{5\times10^{15}}{T_8^6B_{12}^2}(1+5\times10^{-7}L_5^2T_8^8)~{\rm
yr}$ (for the irrotational component), and $t_{\rm
hall}\sim5\times10^{8}\frac{L_5^2T_8^2}{B_{12}}(\frac{\rho}{\rho_{\rm
nuc}})~{\rm yr}$, where $B_{12}$ is the surface magnetic field
strength in units of 10$^{12}$~G, $T_8$ denotes the core temperature
in units of $10^8$~K, $\rho_{\rm
nuc}\equiv2.8\times10^{14}$~g~cm$^{-3}$ is the nuclear density, and
$L_5$ is a characteristic length scale of the flux loops through the
outer core in units of $10^5$~cm.

Assuming $T_8$ does not vary or only varies slowly with time (Pons
2009), the ohmic dissipation produces exponential decay of magnetic
field, $B= B_{0}e^{-t/\tau_{\rm d}}$, where $\tau_{\rm d}$ is
referred to as the decay time scale. The other two mechanisms all
produce power-law decay, $B= B_{0}(t/t_0)^{-\alpha}$, where the
decay index $\alpha=1$ or $\alpha=0.5$ for Hall effect or ambipolar
diffusion (for $T\ll10^{12}$~years), respectively. For the
exponential decay, the true age of a neutron star is given by (Lyne
1975) $T_{\rm e}=\frac{1}{2}\tau_{\rm
d}\log(1+\frac{4}{n-1}\frac{T_{\rm s}}{\tau_{\rm d}})$. For the
power-law decay, the true age of a neutron star is given by $T_{\rm
p}=\frac{2(1-2\alpha)}{n-1}T_{\rm s}$ (for $\alpha<0.5$ and $T_{\rm
p}\gg t_{0}$), or $T_{\rm p}\log(T_{\rm
p}/t_{0})=\frac{2}{n-1}T_{\rm s}$ (for $\alpha=0.5$), or $T_{\rm
p}=\frac{2(2\alpha-1)}{n-1}T_{\rm s}(T_{\rm p}/t_{0})^{2\alpha-1}$
(for $\alpha>0.5$), respectively (in the above $t_0$ and $B_0$ are
the age and dipole magnetic field strength of a neutron star when it
started its current phase of magnetic field decay). Because of the
coupling between $t_{0}$ and $\alpha$ when $\alpha\geq0.5$, we will
first discuss the case $\alpha<0.5$ in the following. We stress that
only $\dot{P}=KP^{2-n}$ and $\dot{n}=0$ are assumed in deriving the
above relations.

In figure~3, we compare predictions of the exponential or power-law
magnetic field decay with the observed correlation between $T_{\rm
s}$ and $T_{\rm SNR}$ of neutron stars; for convenience we take
$n=3$ for the moment and we will discuss more general cases later.
Clearly the exponential decay model cannot describe the data points
for most young neutron stars. The power-law decay model can describe
the data points if different populations of neutron stars are
allowed to take different values of $\alpha$. The dipole magnetic
field strength of each neutron star is shown in the same panel,
suggesting that neutron stars with weaker dipole magnetic field tend
to have larger $\alpha$; this is a generic prediction of the model
if their initial dipole magnetic fields have similar strengths.

\begin{figure}
\centerline{ \hbox{ \epsfxsize=3.0in \epsfbox{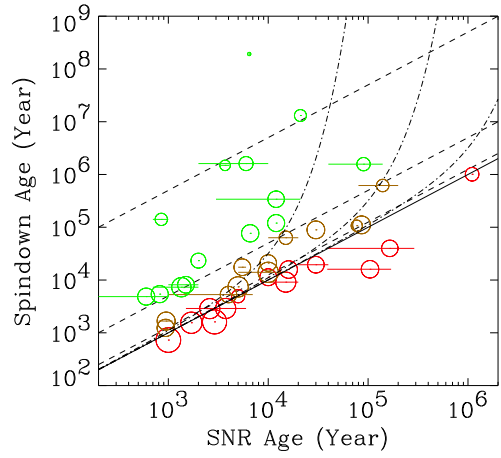}}} \caption{Observed correlation between the
spin-down $T_{\rm s}$ and true ($T_{\rm SNR}$) ages of neutron stars, compared with the predictions of the
exponential or power-law magnetic field decay models. The solid line corresponds to $T_{\rm SNR}=T_{\rm s}$. The
dot-dashed lines correspond to $T_{\rm e}=\frac{1}{2}\tau_{\rm d}\log(1+2\frac{T_{\rm s}}{\tau_{\rm d}})$, with
$\tau_{\rm d}=10^4$, $10^5$, and $10^6$ (in top-down order), respectively. The dashed lines correspond to
$T_{\rm p}=(1-2\alpha)T_{\rm s}$ (for $\alpha<0.5$ and $T_{\rm p}\gg t_{0}$), with $\alpha=0.499$, $0.40$ and
$0.10$ (in top-down order), respectively. The dividing values of $\alpha$ are chosen in such a way that the
numbers of neutron stars in the three regions (in three different colors) are roughly the same, for visual
clarity only. The strength of the dipole magnetic field of each neutron star is also illustrated with the radius
of the circle proportional to $\log(B/(10^{10} (\rm G)))$.}.
\end{figure}

In figure~4, we show the distribution of $\alpha$, derived from
figure~3 by counting the number of neutron stars in each interval of
$\alpha$. A striking bi-modal distribution is obvious, indicating
that about half of neutron stars have no or very weak magnetic field
decay ($\alpha\approx 0$) and the other half have very rapid
magnetic field decay ($\alpha\approx 0.5$). This bi-modal
distribution is actually another manifestation of the bi-modal
distribution shown in figure~1~(upper panel); however here the two
peaks are well separated, indicating that the power-law description
of their dipole magnetic field decay is appropriate and physical. If
we take $\alpha=0.5$, then $T_{\rm p}\log(T_{\rm p}/t_{0})=T_{\rm
s}$, i.e., it is possible to reproduce all observed data points with
$T_{\rm p}/T_{\rm s}>1$ by adjusting $t_{0}$. For $\alpha>0.5$ and
$\alpha-0.5\approx0$, $(T_{\rm p}/t_{0})^{2\alpha-1}\approx 1$ and
thus $\alpha\approx 0.5$ can also reproduce all observed data points
with $T_{\rm p}/T_{\rm s}>1$. Therefore we conclude that the
ambipolar diffusion ($\alpha=0.5$) dominates the magnetic field
decay for nearly half of the young neutron stars associated with
supernova remnants, implying that $t_{\rm ambip}< t_{\rm ohmic}$ and
$t_{\rm ambip}< t_{\rm hall}$. Clearly the first condition can be
easily met for physically plausible parameters of neutron stars. The
second condition, i.e., $t_{\rm ambip}< t_{\rm hall}$ requires
$B_{0}> 6\times 10^{12}$~G, about the dipole magnetic field
strengths of these neutron stars without significant magnetic field
decay, i.e., $T_{\rm SNR}\approx T_{\rm s}$.

\begin{figure}
\centerline{ \hbox{ \epsfxsize=3.0in \epsfbox{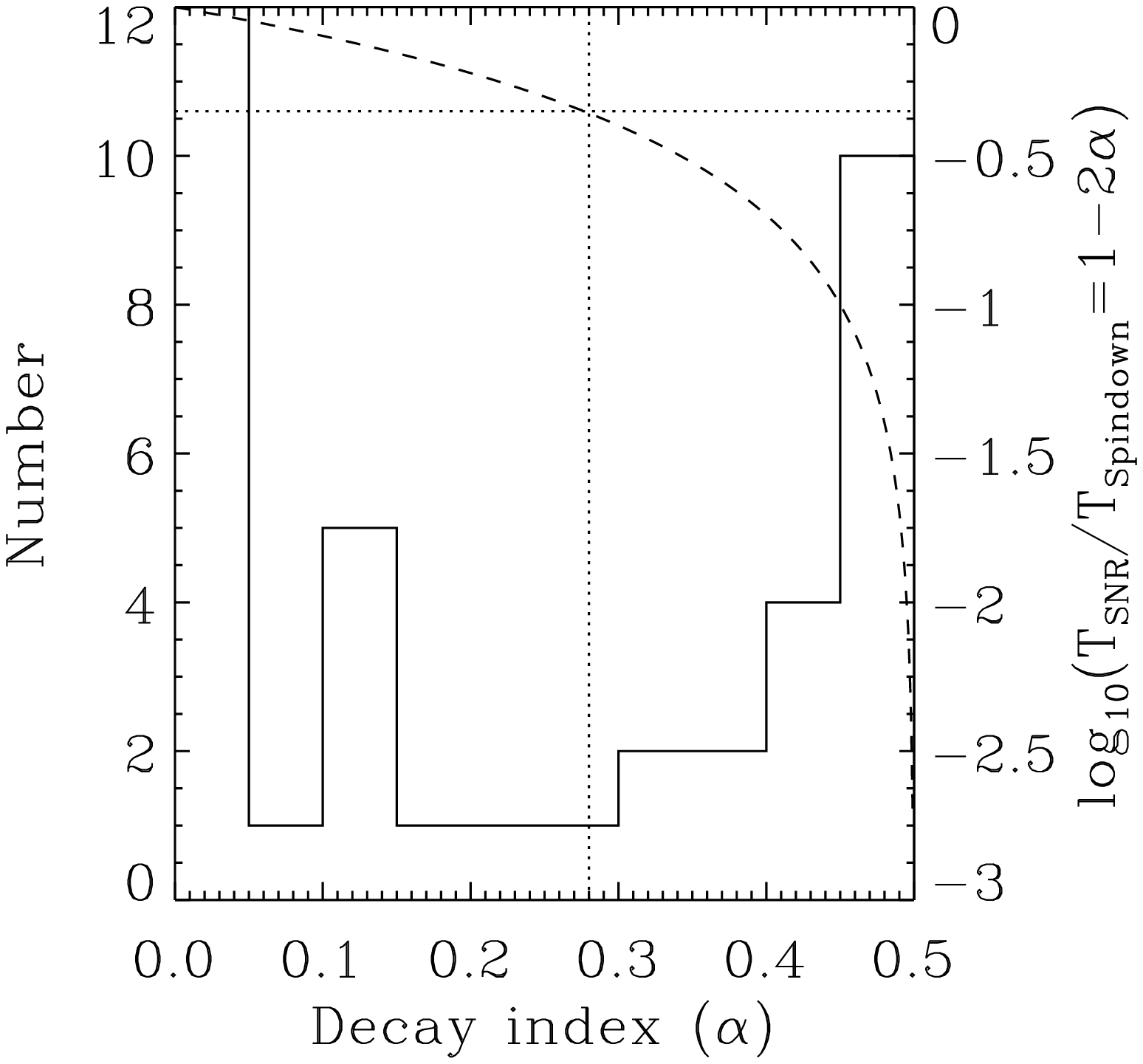}}} \caption{The histogram shows the distribution of
$\alpha$, derived from figure~3 by counting the number of neutron stars in each interval of $\alpha$. The dashed
curve plots $T_{\rm SNR}/T_{\rm s}=(1-2\alpha)$.  The horizontal dotted line corresponds to $T_{\rm
SNR}=0.5T_{\rm s}$. A bi-modal distribution is obvious, indicating that about half of neutron stars have no or
very weak decay ($\alpha\approx 0$) and the other half have very rapid decay ($\alpha\approx 0.5$). This
bi-modal distribution is actually another manifestation of the bi-modal distribution shown in figure~1~(upper
panel).}
\end{figure}

It is notable that there is one neutron star (PSR~B1757--24) with
$T_{\rm s}<T_{\rm SNR}$. In this case, $\dot{K}>0$ is required,
taking $n=3$ as above (please do not confuse $n$ with $n_{\rm b}$
here and throughout this paper). This agrees with the suggestion
that its non-recovered glitches have significantly increased its
dipole magnetic field (Lin 2004). In addition, the distribution peak
at $T_{\rm SNR}=T_{\rm s}$, shown in figure~1~(upper panel),
requires $n=3$, if $\dot{K}=0$ for these neutron stars. Therefore
$n=3$ assumed above is reasonable. On the other hand, the general
expression of the spin-down age is $T_{\rm
s,n}=\frac{1}{n-1}{P/\dot{P}}$. Because $P$ clearly increases with
time statistically, as shown in figure~2~(upper panel), $P\approx
P_0$ is not plausible. Therefore $n\gg3$ is required for these
neutron stars with $T_{\rm SNR}\ll T_{\rm s}$ if there is no
significant magnetic field decay; this is clearly also not
plausible. Alternatively, $n<3$ makes the spin-down age ($T_{\rm
s,n}$) of a neutron star even longer than assuming $n=3$; in this
case more or even all neutron stars have $T_{\rm s,n}>T_{\rm SNR}$,
i.e., $\dot{K}<0$ for more or even all of these neutron stars. We
therefore conclude that for any reasonable values of $n$, at least
some of these neutron stars must have experienced significant decay
of their dipole magnetic fields.

\section{Discussion}

It has been believed that magnetic field decay powers the persistent
X-ray emission of magnetars (Thompson 1995, Thompson 1996). It has
also been suggested that their magnetic field decay plays a
significant role in their thermal evolution (Pons 2009), and thus
has significant implications to understanding the neutron star's
equation of state. Here by comparing the spin-down and true ages of
young neutron stars, we find convincing evidence for magnetic field
decay in about half of them. For magnetic field decay dominated by
the ambipolar diffusion ($\alpha=0.5$), it is easy to show that
$T_8\approx 2.6\times 10^{-5}\frac{B_{12}}{L_5} t^{1/2}$. Taking the
typical values of $B_{12}\sim 1$, $t\sim 5\times10^4$~yr, and
$L_5\sim 1$, its core temperature is predicted to be roughly
$\sim\times10^6$~K, corresponding to a surface temperature (Arras
 2004) of $\sim6\times10^5$~K and consistent with observations (Pons
 2009).

It however remains puzzling why these neutron stars discussed here,
that are observed as normal radio pulsars, have experienced such
diverse evolutionary routes: the magnetic fields increase in very
few neutron stars, remain nearly constant in about half of them, or
decrease rapidly in others. Perhaps these neutron stars were born
with different physical properties (Han 1997); understanding the
initial conditions of neutron stars will inevitably advance our
understanding of stellar evolution, supernova explosion, and the
equation of state of neutron stars significantly.

\acknowledgements  We appreciate discussions with and comments from E. P. J. van den Heuvel, S. Tsuruta, J. L.
Han, W. W. Tian, X. P. Zheng, R. X. Xu, F. J. Lu and H. Tong. SNZ acknowledges partial funding support by the
National Natural Science Foundation of China under grant Nos. 10821061, 10733010, 10725313, and by 973 Program
of China under grant 2009CB824800.


\begin{thebibliography}{}

\bibitem[]{} Arras, P., Cumming, A. \& Thompson, C. 2004, ApJ, 608, L49
\bibitem[]{} Aguilera, D. N., Pons, J. A. \& Miralles, J. A. 2008, ApJ, 673, L167
\bibitem[]{} Bhattacharya, D., Wijers, R. A. M. J., Hartman, J.W., Verbunt, F. 1992, A\&A, 254, 198
\bibitem[]{} Chanmugam, G., Rajasekhar, A., Young, E.J. 1995, MNRAS, 276, L21
\bibitem[]{} Chen, W.C. \& Li, X.D. 2006, A\&A, 450, L1
\bibitem[]{} Geppert, U., Page, D. \& Zannias, T. 1999, A\&A, 345, 847
\bibitem[]{} Goldreich, P., \& Julian, W.H. 1969, ApJ, 157, 869
\bibitem[]{} Goldreich, P. \& Reisenegger, A. 1992, ApJ, 395, 250
\bibitem[]{} Gonthier, P. L. et al. 2002, ApJ, 565, 482
\bibitem[]{} Gunn, J. E. \& Ostriker, J. P. 1970, ApJ, 160, 979
\bibitem[]{} Guseinov, O. H., Ankay, A., Tagieva, S. O. 2004, International Journal of Modern Physics D, 13, 1805
\bibitem[]{} Han, J. L. 1997, A\&A, 489, 485
\bibitem[]{} Harding, A. K. \& Lai, D. 2006, Rep. on Pro. in Phys., 69, 2631
\bibitem[]{} Heyl, J. S. \& Kulkarni, S. R. 1998, ApJ, 506, L61
\bibitem[]{} Holt, S. S. \& Ramaty, 1970, Nature, 228, 351
\bibitem[]{} Lin, J. R. \& Zhang, S. N. 2004, ApJ, 615, L133
\bibitem[]{} Lorimer, D. R. 2011, High-Energy Emission from Pulsars and their Systems. in ¡°High-Energy Emission
from Pulsars and their Systems¡±, Astrophysics and Space Science
Proceedings 21-36 (Springer-Verlag Berlin Heidelberg)
\bibitem[]{} Lyne, A. G. 2004, From Crab Pulsar to Magnetar? Young Neutron Stars and Their Environments,
ed. Fernando Camilo and Bryan M. Gaensler. San Francisco,
(Australia. CA: Astronomical Society of the Pacific)
\bibitem[]{} Lyne, A. G. \& Graham-Smith, F. 2006, Pulsar astronomy, 3rd ed. Cambridge astrophysics series.
(Cambridge, UK: Cambridge University Press)
\bibitem[]{} Lyne, A. G., Ritchings, R. T., \& Smith, F.G. 1975, MNRAS, 171, 579
\bibitem[]{} Ostriker, J. P. \& Gunn, J. E. 1969, ApJ, 157, 1395
\bibitem[]{} Pacini, F. 1969, Nature, 224, 160
\bibitem[]{} Pacini, F., 1968, Nature, 219, 145
\bibitem[]{} Pons, J. A., Miralles, J. A. \& Geppert, U. 2009, A\&A, 496, 207
\bibitem[]{} Popov, S. B. et al. 2010, MNRAS, 401, 2675
\bibitem[]{} Regimbau, T. \& Freitas Pacheco, J.A. 2001, A\&A, 374, 182
\bibitem[]{} Ridley, J. P. \& Lorimer, D.R. 2010, MNRAS, 404, 1081
\bibitem[]{} Ruderman, M. 2005, A Biography of the Magnetic Field of a
Neutron Star, in ¡°The Electromagnetic Spectrum of Neutron Stars¡±,
Proceedings of the 6th NATO ASI series held in Marmaris, Turkey, ed.
A. Baykal, S. K. Yerli, S. C. Inam, and S. Grebenev. (Berlin:
Springer), 47
\bibitem[]{} Thompson, C. \& Duncan, R. C. 1995, MNRAS, 275, 255
\bibitem[]{} Thompson, C. \& Duncan, R. C. 1996, ApJ, 473, 322


\end{thebibliography}

\end{document}